%% file: main_revised.tex
\documentclass[twocolumn]{aastex7}

\usepackage[caption=false]{subfig}

\usepackage{xspace}

\newcommand{\E}{$E\left(B-V\right)$}

\newcommand{\R}{$f_\mathrm{em}/f_\mathrm{abs}$}

\begin{document}

\title{Filling the Pockets: The Spherical Nature of 3D Deflagration in Thermonuclear Supernovae}

\input authors

\submitjournal{ApJ}

\received{\today}
\revised{\today}

\accepted{\today}

\begin{abstract}

We investigate thermonuclear explosions within the delayed detonation framework. While spherical delayed detonation models generally reproduce key observational features, a fundamental inconsistency emerges in three dimensions: 3D hydrodynamic simulations exhibit insufficient white dwarf expansion during the deflagration phase. We identify the early deflagration stage, when the burning is dominated by the laminar speed, as a critical phase and explore potential solutions using three-dimensional magnetohydrodynamic simulations performed with the \texttt{FLASH} code.
 {In absence of preexisting small-scale velocity fields, }
hydrodynamical simulations of the early deflagration phase produce large pockets of unburned C/O, leading to inefficient burning. Much of the released energy is deposited into buoyantly rising plumes rather than into the global preexpansion of the white dwarf, which is required to produce the partially burned layers characteristic of SNe Ia. 
{In contrast, 
when preexisting turbulent velocity fields on scales expected from the smoldering phase are included, the entrainment of burned material into unburned pockets enables the conductive ignition of the surrounding unburned fuel. 
The effective burning approaches that in spherical models, addressing a long-standing problem in multidimensional deflagration models.} {For magnetic fields considered here, $ \lesssim 1\% $ of the saturation strength, we find that the effective burning rate is dominated by the turbulence. Magnetic fields only marginally suppress the rising of burned plumes and the formation of small structures, leading to a slightly more confined burning region and a reduced burning rate.} 

\end{abstract}


\section{Introduction} 
\label{sec:intro}

\input introduction

\section{Magnetic properties} 
\label{sec:magnetic}

\input magnetic

\section{Numerical Setup of the Simulations}
\label{sec:methods}

\input methods

\section{Results}
\label{sec:results}

\input results

\section{Conclusions}
\label{sec:conclusions}

\input conclusions


\begin{acknowledgements}

We thank the anonymous referee for comments that improved the
presentation of our results. Sa.S. and P.H. acknowledge the support by the National Science Foundation NSF awards AST-1715133 and  AST-230639 for enabling the development of the methods and computational codes. P.H. acknowledges support from NASA grants JWST-GO-02114,
JWST-GO-02122, JWST-GO-04522, JWST-GO-04217, JWST-GO-04436,
JWST-GO-03726, JWST-GO-05057, JWST-GO-05290, JWST-GO-06023,
JWST-GO-06677, JWST-GO-06213, JWST-GO-06583, JWST-GO-09231. C.A., K.M., W.B.H. and C.P. acknowledge support from NASA grants JWST-GO-02114, JWST-GO-02122, JWST-GO-03726, JWST-GO-04217, JWST-GO-04436, JWST-GO-04522, JWST-GO-05057, JWST-GO-05290, JWST-GO-06023, JWST-GO-06213, JWST-GO-06583, and JWST-GO-06677. Support for these programs was provided by NASA through a grant from the Space Telescope Science Institute, which is operated by the Association of Universities for Research in Astronomy, Inc., under NASA contract NAS5-03127. W.B.H. acknowledges support from the National Science Foundation Graduate Research Fellowship Program under Grant No. 2236415. The software used in this work was developed in part by the DOE NNSA- and DOE Office of Science-supported Flash Center for Computational Science at the University of Chicago and the University of
Rochester.
This work also required using and integrating a Python package for astronomy, yt (https://yt-project.org, \citealt{Turk2011}).

\end{acknowledgements}

\software{\texttt{FLASH} (\citealt{Fryxelletal2000,Dubeyetal2009,Dubeyetal2013,Dubeyetal2014}, \url{https://flash.rochester.edu}),
\texttt{yt} (\citealt{Turk2011}, \url{https://yt-project.org}),
\texttt{BxC toolkit} (\citealt{Durriveetal2022,Macietal2024}, \url{https://bxc.academy/})}

\bibliography{refs}
\bibliographystyle{aasjournal}

\end{document}

%% file: authors.tex
\newcommand{\FSU}{\affiliation{Department of Physics, Florida State
    University, 77 Chieftan Way, Tallahassee, FL 32306, USA}}
\newcommand{\PSI}{\affiliation{Planetary Science Institute, 1700 East Fort
  Lowell Road, Suite 106,Tucson, AZ 85719-2395 USA}}
\newcommand{\HS}{\affiliation{Hamburger Sternwarte, Gojenbergsweg 112, 21029 Hamburg, Germany}}
\newcommand{\IFA}{\affiliation{Institute for Astronomy, University of Hawai’i at Manoa, 2680 Woodlawn Dr., Hawai’i, HI 96822, USA}}
\newcommand{\VT}{\affiliation{Department of Physics, Virginia Tech,
    850 West Campus  Drive, Blacksburg VA, 24061, USA}}
\newcommand{\GRFP}{\altaffiliation{National Science Foundation Graduate Research Fellow}}
\newcommand{\FINESST}{\altaffiliation{NASA FINESST Future Investigator}}
\newcommand{\NHFPE}{\altaffiliation{NHFP Einstein Fellow}}
\newcommand{\UIUC}{\affiliation{Department of Astronomy, University of Illinois Urbana-Champaign, 1002 West Green Street, Urbana, IL 61801, USA}}
\newcommand{\NSFSIMS}{\affiliation{NSF-Simons AI Institute for the Sky (SkAI), 172 E. Chestnut St., Chicago, IL 60611, USA}}

\newcommand{\STSci}{\affiliation{Space Telescope Science Institute, 3700 San Martin Drive, Baltimore, MD 21218-2410, USA}}
\newcommand{\Carnegie}{\affiliation{Observatories of the Carnegie
    Institution for Science, 813 Santa Barbara St., Pasadena, CA 91101, USA}}
\newcommand{\MSU}{\affiliation{Department of Physics \& Astronomy,
    Michigan State University, East Lansing, MI, USA}}
\newcommand{\TAMU}{\affiliation{George P. and Cynthia Woods Mitchell
    Institute for Fundamental Physics and Astronomy,
    Department of Physics and Astronomy, Texas 
             A\&M University, College Station, TX 77843, USA}}
\newcommand{\IALP}{\affiliation{Instituto de Astrof\'isica de La Plata
    (IALP), CONICET, Paseo del Bosque S/N, B1900FWA La Plata, Argentina}}
\newcommand{\LaPlata}{\affiliation{Facultad de Ciencias Astron\'omicas
    y Geof\'isicas Universidad Nacional de La Plata, Paseo del Bosque,
    B1900FWA, La Plata, Argentina}}
\newcommand{\WPI}{\affiliation{Kavli Institute for the Physics and
    Mathematics of the Universe (WPI), The University of Tokyo,
    Kashiwa, 277-8583 Chiba, Japan}} 

\newcommand{\ICE}{\affiliation{Institute of Space Sciences (ICE,
    CSIC), Campus UAB, Carrer de Can Magrans, s/n, E-08193 Barcelona, Spain}}

\newcommand{\IEEC}{\affiliation{Institut d’Estudis Espacials de
    Catalunya (IEEC), E-08034  Barcelona, Spain}} 

\newcommand{\LCO}{\affiliation{Las Campanas Observatory, Carnegie
    Observatories, Casilla 601, La Serena, Chile}} 

\newcommand{\Aarhus}{\affiliation{Department of Physics and Astronomy,
    Aarhus University, Ny  Munkegade 120, DK-8000 Aarhus C, Denmark.}} 

\newcommand{\OU}{\affiliation{Homer L.~Dodge Department of Physics and
  Astronomy, University of Oklahoma, 440 W. Brooks, Rm 100, Norman, OK
  73019-2061}}  

\newcommand{\UCSC}{\affiliation{Department of Astronomy and Astrophysics,
  University of California, Santa Cruz, CA 95064, USA}} 
\newcommand{\Melbourne}{\affiliation{School of Physics, The University of
  Melbourne, VIC 3010, Australia}}

\newcommand{\LPNHE}{\affiliation{LPNHE, (CNRS/IN2P3, Sorbonne
  Universit\'e, Universit\'e Paris Cit\'e), Laboratoire de Physique
  Nucl\'eaire et de Hautes \'Energies, 75005, Paris, France}}

\newcommand{\Princeton}{\affiliation{Princeton University, 4 Ivy Lane,
    Princeton, NJ 08544, USA}}

\newcommand{\Berkeley}{\affiliation{Department of Astronomy,
    University of California, Berkeley, CA 94720-3411, USA}}

\newcommand{\Tsinghua}{\affiliation{Physics Department, Tsinghua
    University, Beijing, 100084, China}}

\newcommand{\Thailand}{\affiliation{National Astronomical Research
    Institute of Thailand, 260 Moo 4, Donkaew, Maerim, Chiang Mai,
    50180, Thailand}}

\newcommand{\UVA}{\affiliation{Department of Astronomy, University of
    Virginia, 530 McCormick Rd, Charlottesville, VA 22904, USA}}

\newcommand{\LJMU}{\affiliation{Astrophysics Research Institute,
    Liverpool John Moores University, 146 Brownlow Hill, Liverpool L3
    5RF, UK}}

\newcommand{\MPIA}{\affiliation{Max-Planck-Institut f\"ur Astrophysik,
    Karl-Schwarzschild Stra{\ss}e 1, 85748 Garching, Germany}}

\newcommand{\JHU}{\affiliation{Physics and Astronomy Department,
    Johns Hopkins University, Baltimore, MD 21218, USA}}

\newcommand{\OSU}{\affiliation{Department of Astronomy, The Ohio State
    University, Columbus, OH, USA}}

\newcommand{\CCAP}{\affiliation{Center for Cosmology and Astroparticle
    Physics, The Ohio State University, Columbus, OH, USA}}

\newcommand{\MIT}{\affiliation{Department of Physics and Kavli Institute for Astrophysics and Space Research, Massachusetts Institute of Technology, 77 Massachusetts Avenue, Cambridge, MA 02139, USA}}

\newcommand{\CPA}{\affiliation{Centre for mathematical Plasma Astrophysics, Department of Mathematics, KU Leuven, Celestijnenlaan 200B, B-3001 Leuven, Belgium}}

\newcommand{\nextinstitute}{\affiliation{Put the institute of the new author here}}


\author[0000-0001-6107-0887]{S.~Shiber}
\email{sshiber@fsu.edu}
\FSU

\author[0000-0002-4338-6586]{P. Hoeflich}
\email{phoeflich77@gmail.com}
\FSU

\author[0000-0001-5888-2542]{T.~Mera}
\email{tycomera@gmail.com}
\FSU

\author[0009-0001-9148-8421]{E.~Fereidouni}
\email{ef22g@fsu.edu}
\FSU

\author[0009-0000-0636-8322]{Z.~Levy}
\email{zml21@fsu.edu}
\FSU

\author[0009-0000-0636-8322]{D. Maci}
\email{D.macy@gmail.com}
\CPA

\author[0000-0002-5221-7557]{C. Ashall}
\email{cashall@hawaii.edu}
\IFA

\author[0000-0001-5393-1608]{E.~Baron}
\email{ebaron@psi.edu}
\PSI
\HS

\author[0000-0002-9301-5302]{M.~Shahbandeh}
\email{mshahbandeh@stsci.edu}
\STSci

\author[0000-0001-7186-105X]{K. Medler}
\email{kmedler@hawaii.edu}
\IFA

\author[0000-0003-3953-9532]{W.~B.~Hoogendam}
\GRFP
\IFA
\email{willemh@hawaii.edu} 

\author[0000-0002-7305-8321]{C.~M.~Pfeffer}
\GRFP
\IFA
\email{cpfeffer@hawaii.edu}

%% file: introduction.tex
Type Ia supernovae (SNe Ia) are generally agreed to arise from thermonuclear explosions of carbon-oxygen (C/O) white dwarfs (WDs) (e.g., \citealt{hf60}, \citealt{WI73}, \citealt{Piersantietal2004}, and also recent reviews by \citealt{Liu2023}, \citealt{Ruiter2025}, and references therein). Despite much effort and investigation, the community has not yet come to a consensus, neither regarding the progenitor system nor the main mechanism causing the explosion.

In fact, many recent observational and theoretical studies show an intrinsic diversity of SN~Ia \citep{Hoogendametal2025apj,Hoogendametal2025oja,Boseetal2025aa,Boseetal2025arX,Dimitriadisetal2025,Paniaguaetal2026}.
With tuning, most currently favored explosion scenarios can reproduce the overall light curves (LCs) and total flux spectra because nuclear physics governs the progenitor structure, abundances, and explosion energies, a fact described as “stellar amnesia”. 
To truly understand a given event is challenging because the WD can explode in different ways, leading to even more varied burning products, and because it can evolve in very different progenitor systems, resulting in numerous possible interactions between the SN ejecta and components of the progenitor. 

Three leading scenarios are
{\sl I) Explosion of a near-$M_{\rm Ch}$ WD:} 
The flame starts close to the high-density center, producing electron-capture (EC) elements. The resulting structures depend on preexisting turbulence \citep{HoeflichStein2002}, magnetic fields, and develop Rayleigh-Taylor (RT) instabilities. Subsequently, the flame transitions to a detonation  \citep{Khokhlov1991,Khokhlov1995,Niemeyeretal1996,Poludnenkoetal2019,Brookeretal2021,Hristovetal2021}, occurring off-center, leading to asymmetric line profiles and polarization signatures 
with high-density burning probed by stable EC elements. {\sl II) Sub-${\bf {\rm M_{Ch}}}$ explosions:} The explosion is triggered by a surface He-detonation that triggers a second detonation in the C/O core of the WD
\citep{wwt80,Nomoto1982_II,Livne1990,Woosley94,hk96,Kromeretal2010,Sim10,WoosleyKasen2011,Shen2015,Tanikawa2018,Glasneretal2018,Shenetal2018,Townsleyetal2019}. Fundamental characteristics are low-density burning with little production of EC elements, with a significant amount only produced by super-solar metallicity WDs \citep{Blondinetal2022}, and a rather spherical distribution of iron-group elements in the core.
{\sl III) Hydrodynamical, violent, and secular mergers:} Detonations are triggered by collisions, possibly head-on in a triple system, or by compressional heating \citep{Webbink1984,IbenTutukov1984,benz90,rasio94,hk96,segretain97,Yoonetal2007,WMC09,WCMH09,loren09,Pakmor10,isern11,Pakmoretal12,Rosswog2009,Thompson2011,Pejchaetal2013,Kushniretal2013,Dongetal2015,Garcia-BerroHB2017}. The merger can also take place within the common envelope of an
asymptotic giant branch (AGB) star and a WD \citep{HoeflichKhokhlov1996,Yoonetal2007,KashiSoker2011,HoeflichHB2017}. Simulations show ejecta with large-scale rotational symmetry in the density structure that lead to continuum polarization \citep{Patatetal2012}, and spiral structures on small scales in iron-group elements with little EC elements. However, in violent and secular mergers of two WDs, with $M_{\rm WD}\geq 0.7 M_{\odot}$ each, high-density burning and EC elements are expected because it will form hydrostatic WD stabilized by rotation, shrinking by subsequent loss of angular momentum.

Recently, the ubiquity of EC elements has been well established \citep{Galbanyetal2019,Blondinetal2022,DeKacyetal2023,DerKacyetal2024,Ashalletal2024,Kumaretal2025,Kwoketal2025}, suggesting high-density burning.
This favors near-$M_{{\rm Ch}}$ scenarios or secular mergers. However, low continuum polarization values are observed in the early phase of many SNe, disfavoring the merger scenario. Spectropolarimetry confirmed an overall small deviation from sphericity, but also pointed
toward a large-scale asymmetry in the abundance structure, suggesting that the transition from deflagration to detonation occurs off-center \citep{Yangetal2020,Patraetal2022,Hoeflichetal2023,Cikotaetal2026}.

Though spherical delayed-detonation models \citep{Khokhlov1991} have been shown to agree with the lightcurve and the optical to mid-IR spectra \citep{Hoeflichetal2017, DeKacyetal2023,Ashalletal2024,DerKacyetal2024}, the ignition and propagation of the flame are inherently multidimensional. Turbulence, deflagration–detonation transitions, and incomplete burning introduce asphericities that influence nucleosynthesis yields and emergent spectra.
In spherical geometry, the preexpansion required is consistent with observations.
However, three-dimensional (3D) simulations produce large-scale instabilities, and most of the 
energy goes into the rise of burned plumes \citep{Khokhlov1991,Gamezoetal2003,Roepkeetal2003} rather than preexpansion. The ratio between burned and unburned material remains small.

The key question to be addressed is the influence of preexisting  turbulence produced during
the smoldering phase prior to the explosion, and the influence of magnetic fields on the early deflagration.
To address these questions, we present the results of full 3D simulations, which include
both high-turbulent fields and $B$ fields. In Section \ref{sec:magnetic}, we focus on the magnetic properties present in WDs at the beginning of the deflagration phase. In Section~\ref{sec:methods}, we describe our numerical methods. The results are discussed and conclusions are presented in Sections~\ref{sec:results}~and~\ref{sec:conclusions}.

%% file: magnetic.tex
Typically, low magnetic fields are found in WDs \citep{liebert03}. However, observations
of late-time LCs, line profiles, and spectra suggest the existence of high-magnetic fields in excess of $\approx 10^{6}-10^{7}~{\rm G}$ (e.g., \citealt{Diamondetal2018,Hristovetal2021}). A seed magnetic field can be amplified by a dynamo operating in the convective zone or by large-scale circulation in WDs \citep{parker79,Thomas95,Brandenburg05}.
{In addition, plasma instabilities can generate a strong magnetic field even without the presence of a seed magnetic field \citep{Desaietal2025}}.

In large-scale dynamos, a toroidal field is produced by winding up the poloidal component. The convective elements move upward and downward, perpendicular to the toroidal field, creating a new poloidal component 
\citep{parker79}. Alternatively, in small-scale dynamos, convection alone can produce a small-scale 
unstructured magnetic field 
\citep{Brandenburg05,Beresnyak12,Tayler73,Acheson78,Hawley96,Spruit02,braithwaite09,Duez10a,Duez10c}. 
Within dynamo theory and ideal magnetohydrodynamics (MHD), the maximum size of the $B$ field in WDs may approach saturation strength of $10^9-10^{11}$ G in WDs 
\citep{Chandrasekhar56a,Chandrasekhar56b,Mestel56}.
Large-scale dynamos grow with a typical time scale of the Alfv\'{e}n time, 
$t_A\approx R(4\pi\rho)^{1/2}B^{-1}\approx 300$ s \citep{parker79}, while 
small-scale dynamos grow from small scale to large with a timescale of 
$\approx Lv^{-1}\approx1$ s, where $L$ and $v$ are the characteristic scale and
velocity.

In short, the amplification of the magnetic field is driven by either of the following dynamo mechanisms: I) due to accretion over many years; II) due to turbulent burning prior to the runaway, i.e., during the smoldering phase \citep{HoeflichStein2002}; III) due to RT instabilities and magnetic entanglement during the deflagration phase \citep{Hristovetal2018}; 
or IV) due to instabilities induced by $^{56}{\rm Ni}$-decay over time-scales of several days after the explosion (see \citealt{Fesenetal2018} for supporting evidence of caustic structure in SN1885). {For WD mergers, in addition, the amplification of the magnetic field can be driven either by a small-scale dynamo, before and during the merger process, or by a large-scale dynamo, several rotation periods past the merger \citep{Dayaletal2013,Briggsetal2018,Pakmor2024}}.

%% file: methods.tex
We use the multiphysics code \texttt{FLASH} \citep{Fryxelletal2000}, version 4.8, to carry out 3D MHD simulations of centrally ignited deflagration front embedded within turbulent velocity and magnetic fields in a near-$M_{\rm Ch}$ WD of {$M_{\rm WD}=1.35 M_\odot$}. The {nonrotating} WD structure originated from a $7~M_{\rm \odot}$ zero-age-main-sequence star of solar metallicity. {Similarly to what has been found by different groups, the core is carbon depleted as a result of stellar central helium burning \citep{1989ApJ...344..239C,1999ARA&A..37..239B,2001ApJ...557..279D}. In these studies, the $^{12}{\rm C}(\alpha,\gamma)^{16}{\rm O}$ rate and the chemical mixing have been calibrated according to globular clusters or the sun. Moreover, similar structures are found in seismology studies of WDs \citep{1994ApJ...430..839W,2002ApJ...573..803M,2003ASIB..105..251M}. 
}

{After evolving to a WD, it} accretes hydrogen (and helium) from a companion 
\citep{2001ApJ...557..279D,Hoeflichetal2017}. The accretion rate is adjusted to lead to a thermodynamic runaway at a central density of $\approx 10^9~{\rm g~cm^{-3}}$. {This phase of pycnonuclear
burning prior to the thermodynamic runaway is followed using the description introduced by, e.g. \cite{SugimotoNomoto1980,Nomoto1982_I}, with an equation of state for partially degenerate and partially
relativistic Bose- and Fermi-gas plus interactions due to effects of Coulomb corrections, quantum-relativistic effects
on the electron component, and electron-positron pair production and electron screening (for more details, see \citealt{Hoeflich_2002,2019nuco.conf..187H,Hristovetal2021}, and references therein)}. 
Due to accretion, the WD contracts to a radius of $\approx 2,400~{\rm km}$.

We map the WD structure to the 3D Cartesian cubical grid of \texttt{FLASH}. 
To accommodate turbulence of varying scale sizes, our simulations differ in resolution. The base resolution in all of our simulations consists of $\left[32\times32\times32\right]$ cells, which corresponds to \texttt{FLASH}'s three levels of refinement, while the maximal refinement level $l_{\rm max}$ spans from 7 to 10. The grid is logarithmically refined from the center out, while ensuring that the innermost region is refined to the maximal level. 
With a domain size of $5\times10^3~{\rm km}$, this corresponds to a minimal cell size of $dx_{\rm min}=10~{\rm km}$, $2.5~{\rm km}$, and $1.2~{\rm km}$ for $l_{\rm max}=7,~9,~{\rm and}~10$, respectively.

The gravity is computed according to \texttt{FLASH}'s Poisson multipole solver with a highest multipole order of 16 and isolated boundary conditions. A Helmholtz equation of state (EoS) for gas with degenerate electrons is used \citep{TimmesSwesty2000}. This EoS includes contributions to energy and pressure from degenerate electrons–positrons, thermal ions, radiation, and Columb corrections. 

The deflagration is modeled according to the advection-diffusion-reaction (ADR) scheme \citep{Khokhlov1995} with a sharpened Kolmogorov-Petrovski-Piskunov (sKPP) reaction term \citep{Vladimirova2006}, where default values for $\epsilon_0,~\epsilon_1,~f$,~and~$b$ were used (see \citealt{Townsley2007}). Since we focus on the early phase of the deflagration, where the front remains in regions with densities $>10^{7}{\rm ~g~cm^{-3}}$, we assume nuclear statistical equilibrium (NSE) burning, which is tracked by a scalar field $\phi$ (zero means unburnt C/O mixture, while one means full NSE burning to iron-group elements of mostly $^{56}{\rm Ni}$).

{For simplification, a constant laminar flame speed  of $200 ~{\rm km/sec}$ has been adopted. Note that the flame speed equals the conduction speed, which increases with density, namely $\propto \sqrt{P}$ in nonturbulent media \citep{TimmesWoosley1992}. Here, a velocity somewhat higher than the conductive speed has been adopted to mimic the small, nonresolved scales of the preexisting turbulent field and the flame-induced turbulence later on. In the literature, the laminar flame speed is
often taken as the maximum of the conduction speed and the turbulent speed \citep{khokh01,Gamezoetal2003}, which is comparable to our value of the initial turbulent velocity.} 

The nuclear (specific) energy deposition rate is: $\dot{E}_{\rm nuc}=\left(\dot{\phi} /m_p\right)\times\left[\tilde{b}_{\rm Ni} - (1 - {\rm X}_{\rm C})\tilde{b}_{\rm O} - {\rm X}_{\rm C}\tilde{b}_{\rm C}\right]$, where $\tilde{b}_{\rm Ni}$, $\tilde{b}_{\rm O}$, and $\tilde{b}_{\rm C}$ are the binding energies per nucleon of $^{56}{\rm Ni}$, $^{16}{\rm O}$, and $^{12}{\rm C}$, respectively. In the WD model, the core is depleted of carbon and ${\rm X}_{\rm C}=0.278$ (up to a radius of $\lesssim 900~{\rm km}$). Therefore, we assume this composition for the entire simulation evolution and set the energy deposition rate to $\dot{E}_{\rm nuc}= 7.19\times10^{17} \dot{\phi}~{\rm erg~g^{-1}}$.
In addition, we do not change the composition according to the burning, which might have a small effect on the temperature. 

For the initialization, a fully developed turbulent field has been constructed using the BxC toolkit. The BxC toolkit is designed to generate fully customizable synthetic turbulent 3D magnetic fields that include both turbulent structures and higher-order statistics \citep{Durriveetal2022,Macietal2024}. A turbulence with velocities up to 500 km/sec, resulting in a size of the turbulent eddies of up to $\approx 50$ km, has been shown to form during the smoldering phase (see Fig. 8 of \citealt{HoeflichStein2002}). We adjust the turbulence properties to closely follow these physical conditions, i.e., we set the diffusion radius of the Kolmogorov spectrum to $\approx 50$ km and the $v_{{\rm rms}} $ up to $\approx 200 $ km/sec \footnote{{For demonstration of the macroscopic effect of initial turbulent speed, $v_{\rm rms}$ has been reduced in some of the simulations, though the laminar speed has been kept the same.}}
.

\cite{Hristovetal2021} found that an initial large-scale dipole magnetic field has become turbulent during the deflagration phase {because the hydro time-scales are shorter than the amplification time-scales of $B$.} Consequently, it can be expected that the morphology of the magnetic field would become turbulent even before the deflagration starts, and a turbulent magnetic field would develop during the smoldering phase before the runaway. Therefore, we assumed the same morphology of the initial $B$ field as the turbulence. The magnetic field strength can be larger than the minimal values of $10^{6}-10^{7}~{\rm G}$ suggested by observations. The amplification depends on the dynamo time-scales prior to the runaway, and can grow up to the limit of equilibrium between the magnetic field and plasma (equipartition field). Here, we consider fields up to $10^{12}~{\rm G}$, {less than a percent} of the equipartition field \footnote{{Note that the initial turbulent field may not resemble hydrodynamical turbulence if the magnetic pressure becomes comparable to the gas pressure or the amplification of $B$ occurs on shorter timescales, e.g., during the phase of the thermonuclear runaway.}}.

To solve the MHD equations, we use \texttt{FLASH}'s unsplit staggered mesh solver, which ensures the divergence-free constraint of the magnetic fields by default. We assume ideal MHD, and use a hybrid type of Riemann solver, combining both the Roe solver for high accuracy and HLLD for stability. {The MHD boundary conditions assume a zero velocity gradient.} 

{For stability, after mapping the initial WD structure from the spherical to the Cartesian grid, we damp the velocity field for $0.1~{\rm s}$ before igniting. In addition, to avoid inflow from the surrounding low-density medium, we zero the velocity outside a given radius of $1,500~{\rm km}$ from the center at every timestep.}

\begin{figure*}
\centering  
    \includegraphics[scale=0.55, trim={0 6.2cm 0 6.7cm}, clip]{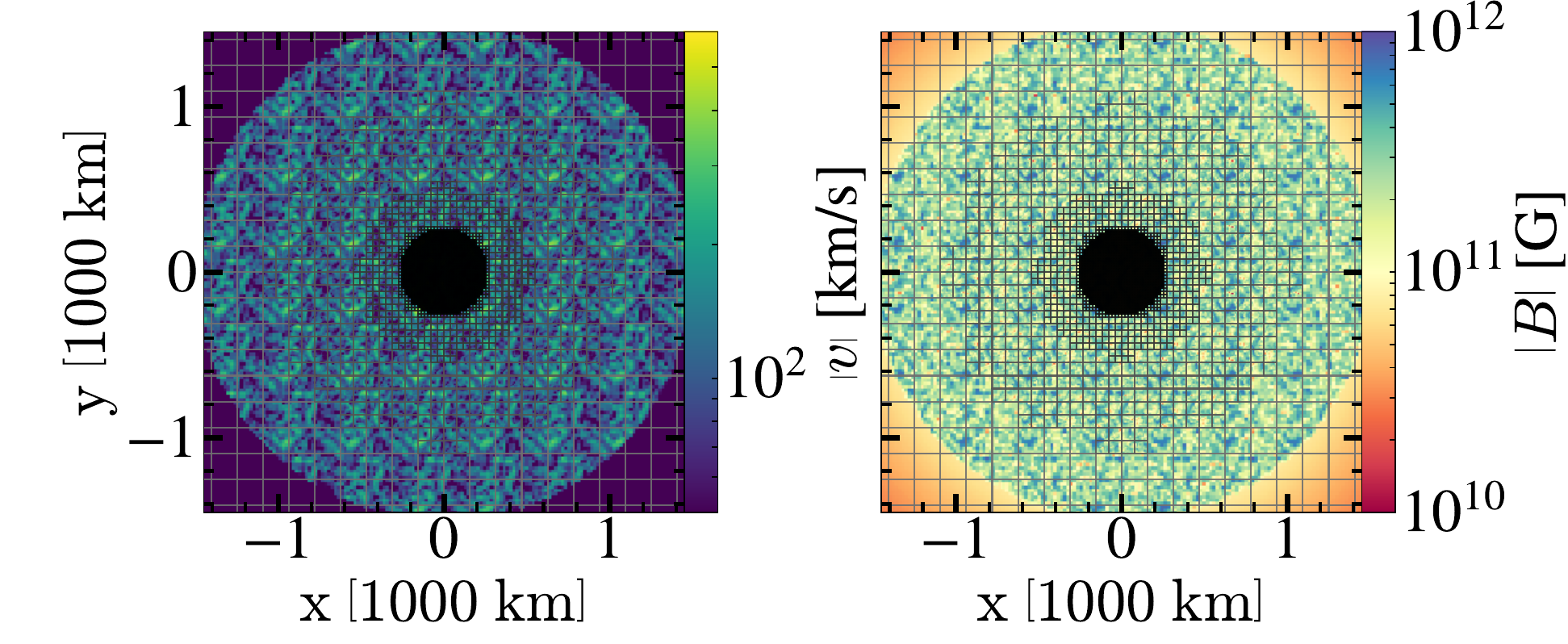}
    \includegraphics[scale=0.55, trim={0 6.2cm 0 6.7cm}, clip]{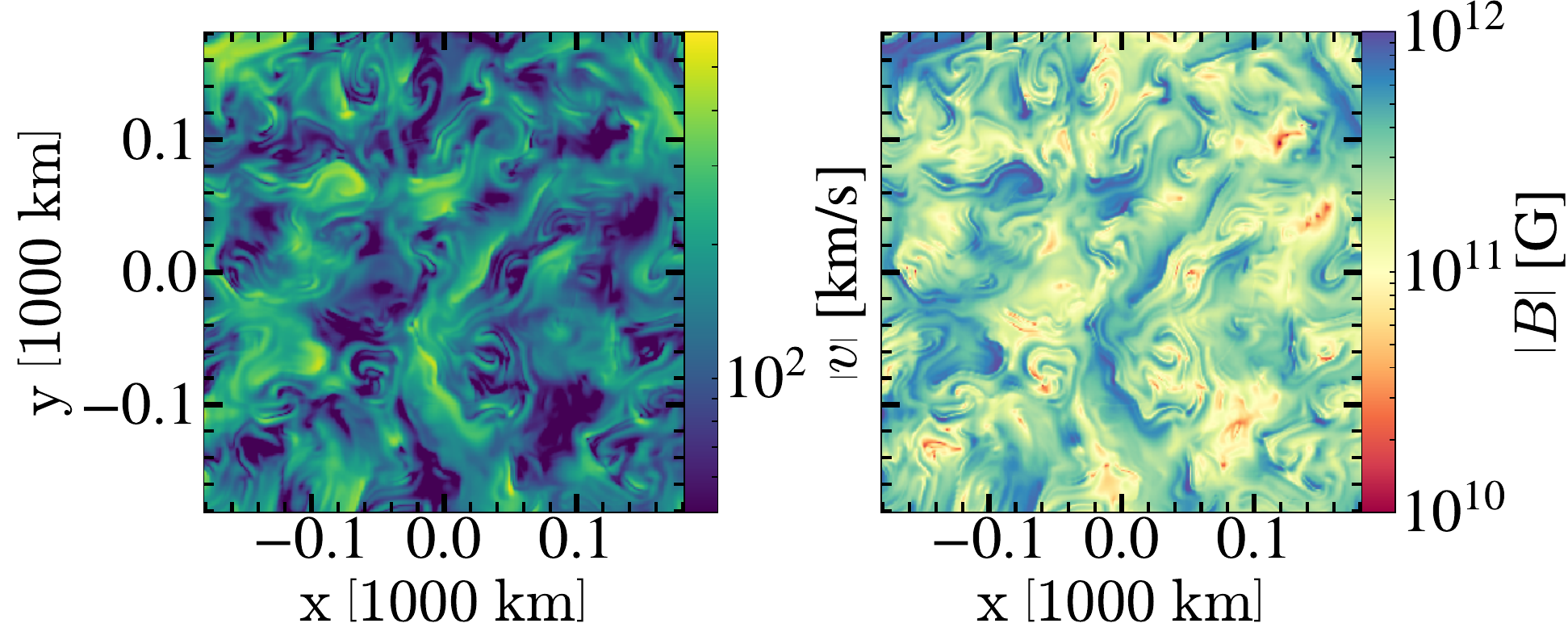}
    \caption{ The initial turbulent velocity (left) and magnetic (right) fields in the highest resolution smallest-scale turbulence simulation D40V170B12, corresponding to smallest eddies size of roughly $40~{\rm km}$ (Kolmogorov's diffusion scale), an rms velocity of $170~{\rm km~s^{-1}}$ and an rms magnetic field of $3.4\times10^{11}~{\rm G}$. The upper row presents the full grid, while the lower row presents a zoomed-in view of the central most refined region.}
    \label{fig:turb_setup}
\end{figure*}

In Figure~\ref{fig:turb_setup}, we show the initial turbulent conditions in the equatorial, xy, plane of our smallest-scale turbulence simulation, D40V170B12. On the left panels, we present the velocity magnitude, while the magnetic field magnitude is presented on the right. In the upper panels, a large-scale view is shown with a representation of our adaptive grid, while in the lower panels, we zoom into the most refined 300 km central regions. The $\approx 50~{\rm km}$ eddies are clearly shown in the lower panels. For this simulation, $v_{\rm rms}=170~{\rm km/s}$ and $B_{\rm rms} = 3.4\times10^{11}~{\rm G}$. Outside of this most inner region, the turbulent pattern is duplicated.
The properties of our simulations are summarized in Table~\ref{tab:prop}. 
\begin{table*}
\centering
\begin{tabular}{l|llllll|rrrrr}
Simulation & $r_D$ & $v_{\rm rms}$ & $B_{\rm rms}$ & $l_{\rm max}$  & $E_{\rm kin}^i$ & $E_{\rm mag}^i$ & $r_{\rm max}^{t=0.5}$ & $f_{300}^{t=0.5}$ &  $f_{500}^{t=0.5}$ & $E_{\rm nuc}^{t=0.5}$ & $E_{\rm kin}^{t=0.5}$ \\
 & (km) & (km/s) & (G) &  & (erg) & (erg) & (km) & & & (erg) & (erg) \\
\hline
D300V30B0  & 300 & 30  & 0.0                & 7  & $1.9\times10^{46}$ & $0$                &
400  & 0.11 & 0.02 & $1.7\times10^{49}$ & $3.2\times10^{47}$ \\
D80V170B12 & 80  & 170 & $3.4\times10^{11}$ & 9  & $2.4\times10^{47}$ & $5.8\times10^{46}$ & 
830  & 0.50 & 0.24 & $6.8\times10^{49}$ & $3.6\times10^{48}$ \\ 
D40V170B12 & 40  & 170 & $3.4\times10^{11}$ & 10 & $2.4\times10^{47}$ & $5.7\times10^{46}$ &
990 & 0.85 & 0.59 & $1.3\times10^{50}$ & $8.5\times10^{48}$ \\
D40V50B9   & 40  & 50  & $3.4\times10^{8}$  & 10 & $2.3\times10^{46}$ & $5.8\times10^{40}$ &
750  & 0.71 & 0.30 & $6.5\times10^{49}$ & $2.4\times10^{48}$ \\
D40V170B6   & 40  & 170  & $3.4\times10^{5}$  & 10 & $2.4\times10^{47}$ & $5.7\times10^{34}$ &
1000  & 0.88 & 0.62 & $1.4\times10^{50}$ & $9.0\times10^{48}$ \\
D40V20B12   & 40  & 20  & $3.4\times10^{11}$  & 10 & $4.0\times10^{45}$ & $5.8\times10^{46}$ &
590  & 0.66 & 0.18 & $4.1\times10^{49}$ & $1.0\times10^{48}$ \\

\end{tabular}
\caption{Properties of our simulations. The binding energy of the WD (sum of gravitational potential and internal energy) is $E_{\rm bind} = 4.8\times10^{50}~{\rm erg}$. Each simulation is given a name based on their smallest scale diffusion radius ($r_D$) of the turbulence, DXXX, follows by the turbulence velocity rms ($v_{\rm rms}$), VXXX, and $\log_{10}$ of the magnetic field rms ($B_{\rm rms}$), BX (note though that B0 is a pure hydrodynamic simulation). The maximal level of refinement $l_{\rm max}$, the initial kinetic and magnetic energy are additionally shown (columns 3-5). The rightmost five columns show deflagration status at $t=0.5~{\rm s}$, when we stopped our simulations. $r_{\max}$ is the maximal distance of the deflagration front from the center, $f_{300}$ and $f_{500}$ are the fractions of burnt mass inside a sphere of 300 and 500 km, respectively. $E_{\rm nuc}$ and $E_{\rm kin}$ are the total deposited nuclear energy and kinetic energy, respectively.}
\label{tab:prop}
\end{table*}

%% file: results.tex
We initiate the simulations by igniting the most central $100~{\rm km}$ of the WD. We evolve the simulations until $t=0.5~{\rm s}$ and follow the propagation of the deflagration to study the effect of preexisting turbulent magnetic and velocity fields on the efficiency of burning. We stopped at $t=0.5~{\rm s}$ because in this study we focus only on the early evolution. By this time, the maximum distance that the deflagration front has reached in our simulations is $\lesssim 1000~{\rm km} $. {This is $\lesssim 2/3$ than the radius, outside of which we zero the velocities ($1,500~{\rm km}$), and therefore zeroing the velocities outside this radius has only a marginal effect on our results.}

\begin{figure*}
\centering  
    \includegraphics[scale=0.55, trim={0 4.2cm 0 0.5cm}, clip]{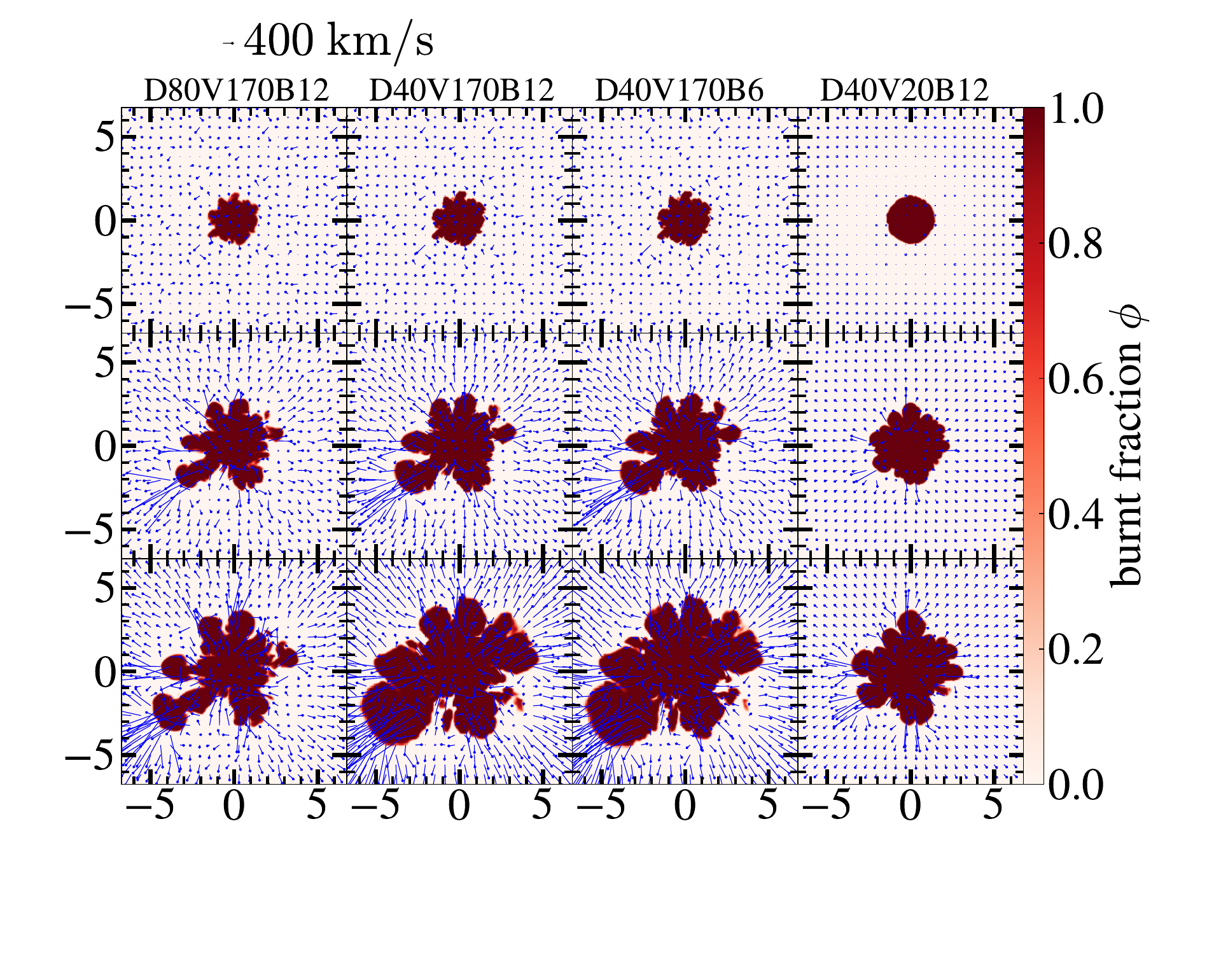} 
    \caption{Fraction of burnt material at three times, $t=0.15,~0.35,~0.45~{\rm s}$ (top, middle, bottom panels, respectively), at the equatorial plane comparing our simulations. The axes are in units of $100~{\rm km}$. From left to right, the simulations shown are:  {D80V170B12, D40V170B12, D40V170B6, and D40V20B12} as denoted at the top of each column. Velocity vectors are plotted as blue arrows.}
    \label{fig:comp_models_flam_xy}
\end{figure*}

\begin{figure*}
\centering  
    \includegraphics[scale=0.55, trim={0 4.8cm 0 4.5cm}, clip]{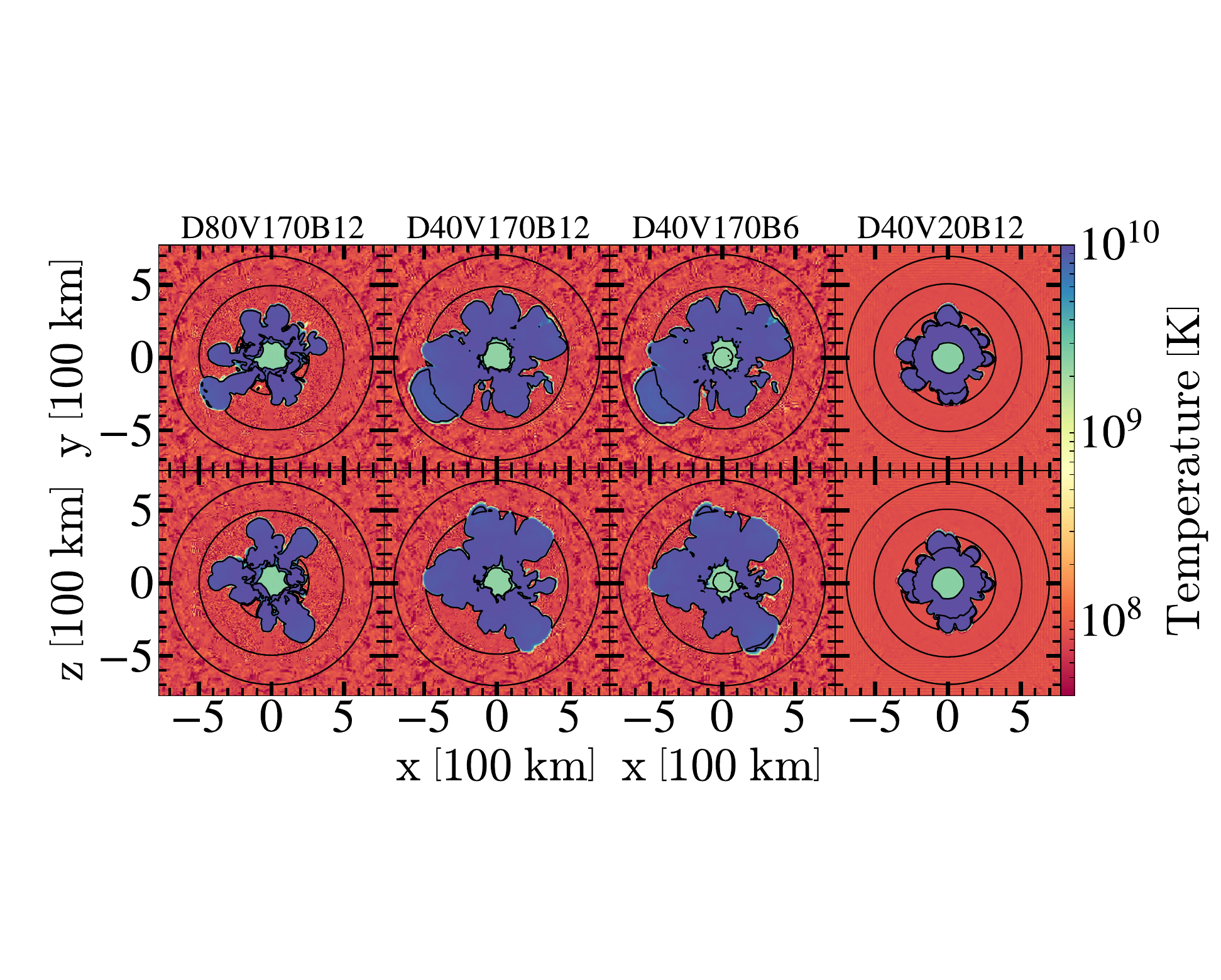}  
    \caption{{Temperature slices at $t=0.45~{\rm s}$ of the equatorial (top row) and meridional (bottom row) planes, comparing four selected simulations with different magnetic field strengths, turbulent conditions, and turbulent sizes. The contours (black) correspond to densities of $4,~6,~8,~10\times10^{8}{\rm~g~cm^{-3}}$. As default, a turbulent scale of 40 km is used, consistent with hydrosimulations for the thermonuclear runaway \citep{HoeflichStein2002}. Though the turbulent field controls the burning (column 2 vs 3), the magnetic field
    constrains the rise of the plume to higher density regions. $B$ partially suppresses the small-scale structure. Note that turbulent scales larger than the typical size of the pockets are ineffective in mixing burned material into unburned pockets, but rather result in an overall more fragmented morphology (column 1 vs 2).   
    }}
    \label{fig:comp_models_t045_temp}
\end{figure*}

In Figure~\ref{fig:comp_models_flam_xy}, we show the evolution of the deflagration in our simulations. In each panel, we plot the fraction of burned material, $\phi$, in the equatorial plane. 
{In Figure~\ref{fig:comp_models_t045_temp}, we show temperature slices both along the equatorial and meridional planes, but only at $t=0.45~{\rm s}$.}

The deflagration front starts to propagate with the laminar speed (Figure~\ref{fig:comp_models_flam_xy}~upper~row). 
Subsequently, RT instabilities begin to develop, creating plumes of burned material (middle row) that rapidly move outward due to buoyancy forces (lower row). These rising plumes can leave large pockets of unburnt material between them, which cannot be reached by the laminar front {right away. The size of these pockets can be approximated as the largest distance between two adjacent rising plumes. For the hydrodynamic simulation with a large-scale turbulent field (D80V170B12; first column from left), we find a pocket size of $\approx 100~{\rm km}$, which is very similar to the pure hydrodynamical model (D300V30B0).}

However, {small-scale  turbulence drags the burned material into these pockets through passive flow and increases the burning. } 
{In principle,} Lorentz force acts perpendicularly to the motion of the rising plumes, curving the deflagration front, and consequently filling the pockets more efficiently. {For $B$ much smaller than the equipartition field as considered here, this effect is minor, and the main effect is burning away the smallest structures (comparing the second and third columns in Figures~\ref{fig:comp_models_flam_xy}~and~\ref{fig:comp_models_t045_temp}).}

The key factor is the size of the turbulence smallest scale (i.e., the diffusion radius). This is most notable when comparing between simulations { D80V170B12 and D40V170B12 (first and second columns)}. These simulations are identical in their turbulence strength and magnetic field strength (see their initial kinetic and magnetic energies in Table~\ref{tab:prop}), and only differ in their turbulence's diffusion radius. The small-scale turbulence in D40V170B12 bends the deflagration front more efficiently, allowing the laminar speed to consume the unburned matter. The general result is a reduction in the distance between {plumes of} the burned material. {The burning within a radius is more complete and resembles a spherical deflagration both in the xy and xz planes (Figure~\ref{fig:comp_models_t045_temp} upper and lower rows, respectively) .}

\begin{figure*}[!ht]
\centering  
    \subfloat[D300V30B0]{\includegraphics[scale=0.36]{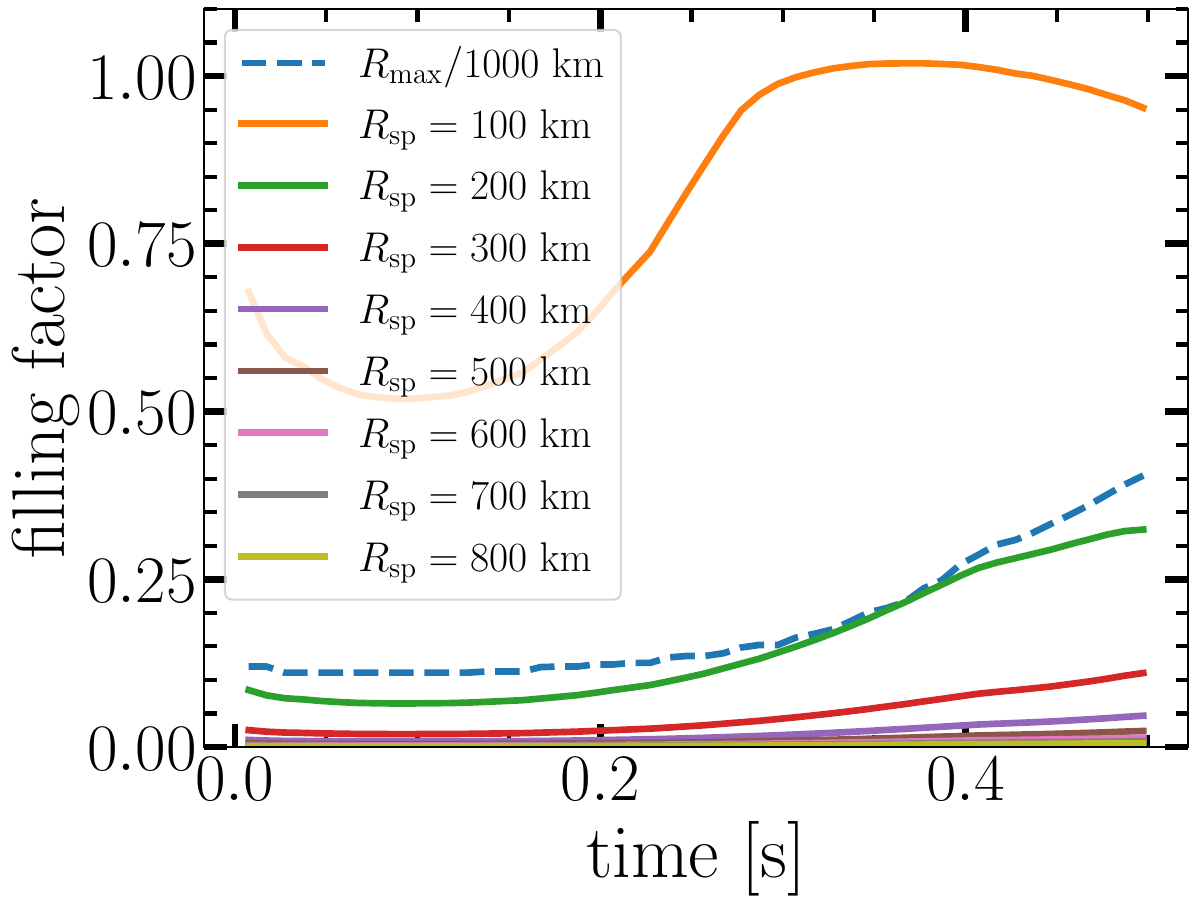}}   
    \subfloat[D80V170B12]{\includegraphics[scale=0.36]{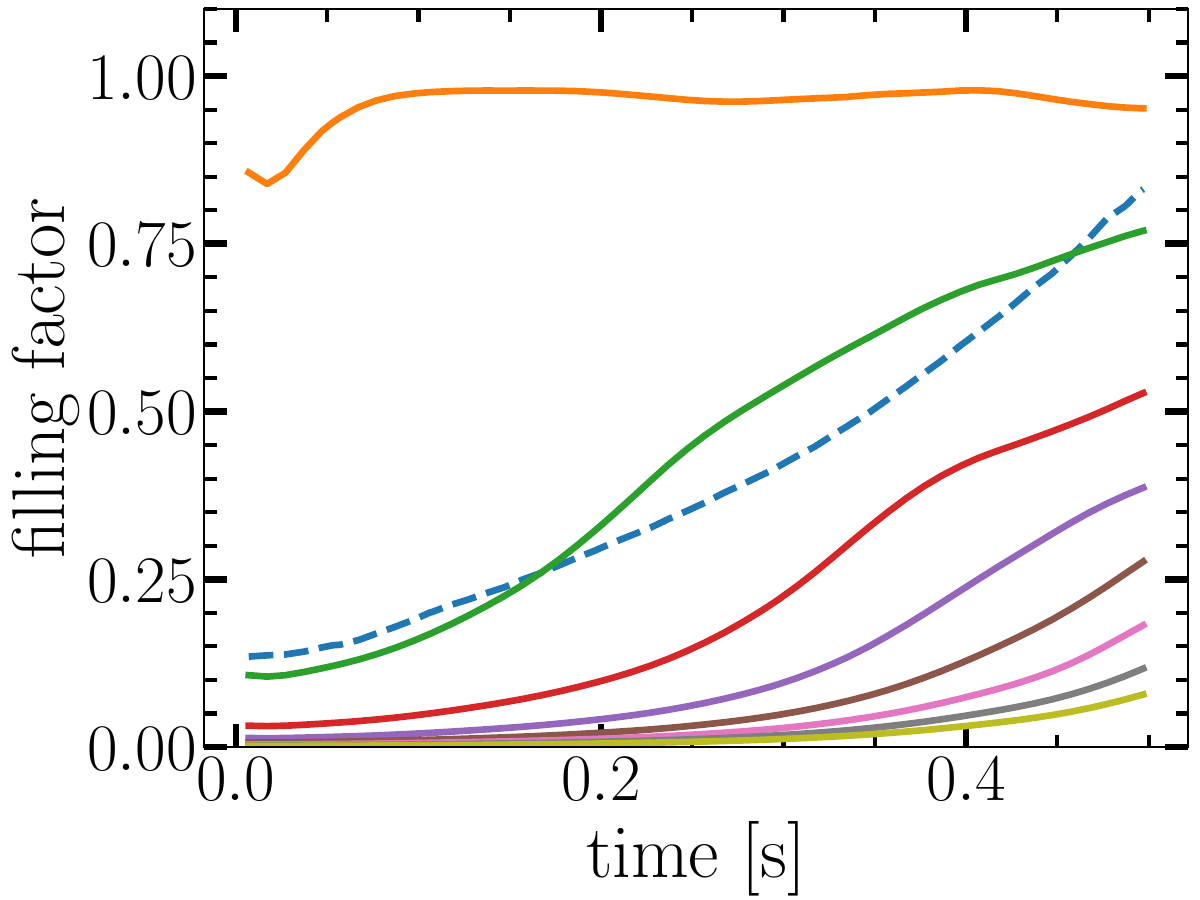}}
    \qquad
    \subfloat[D40V50B9]{\includegraphics[scale=0.36]{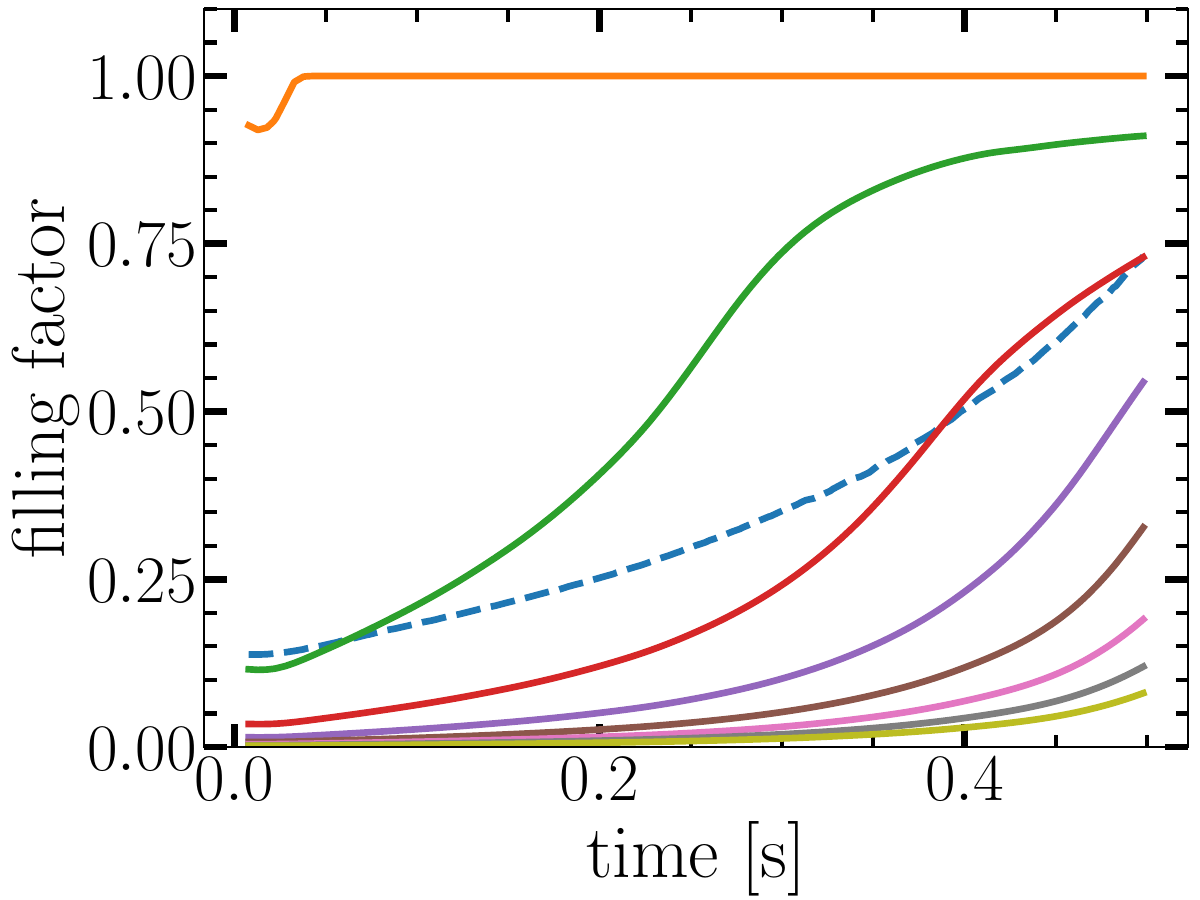}}
    \subfloat[D40V170B12]{\includegraphics[scale=0.36]{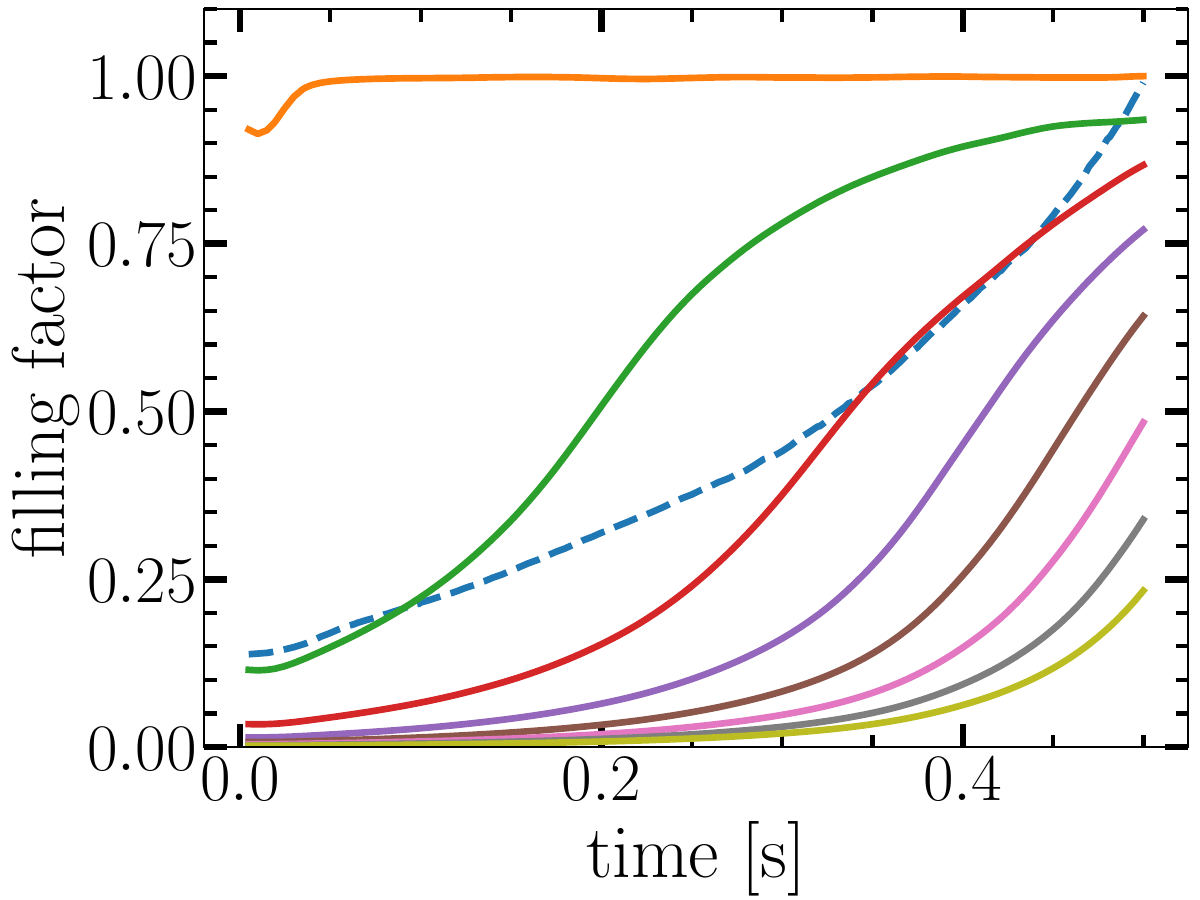}}
    \qquad
    \subfloat[D40V20B12]{\includegraphics[scale=0.36]{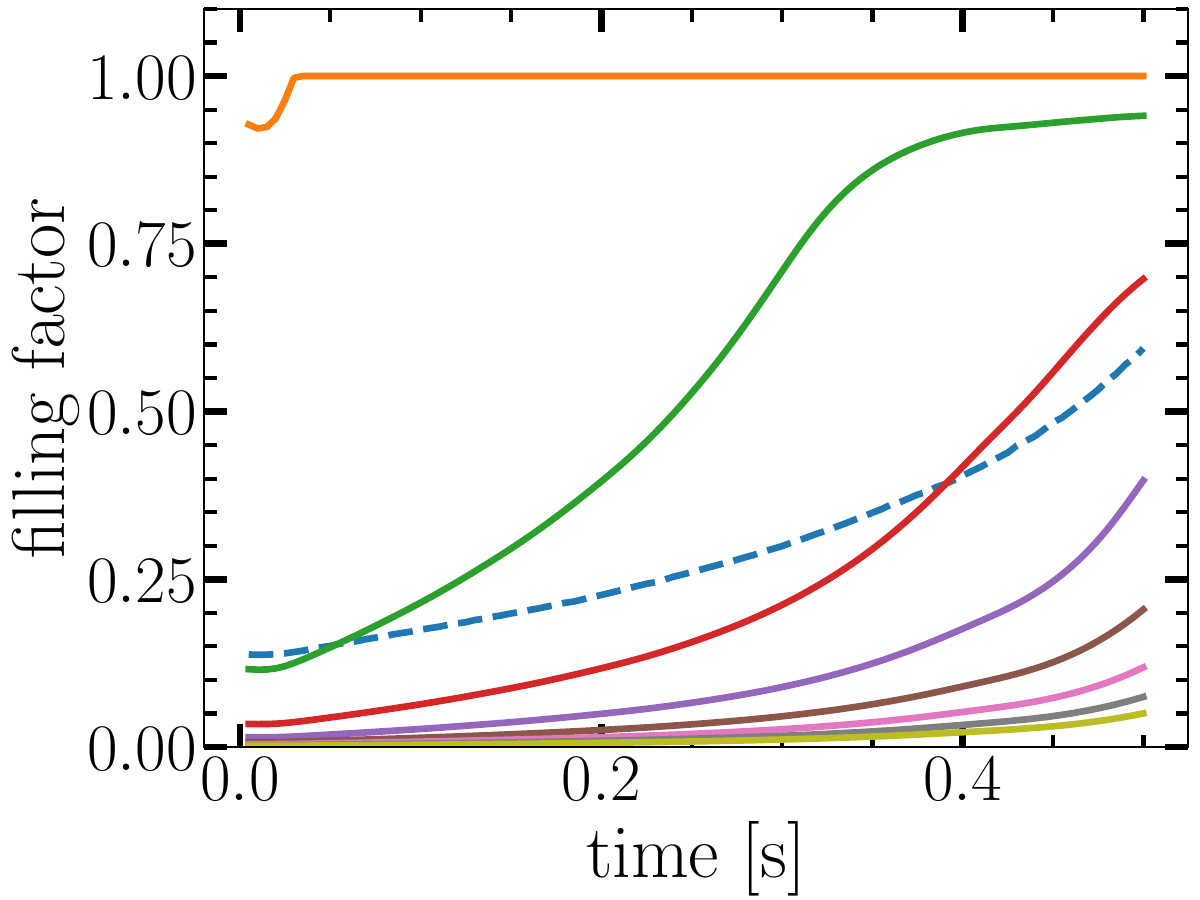}}
    \subfloat[D40V170B6]{\includegraphics[scale=0.36]{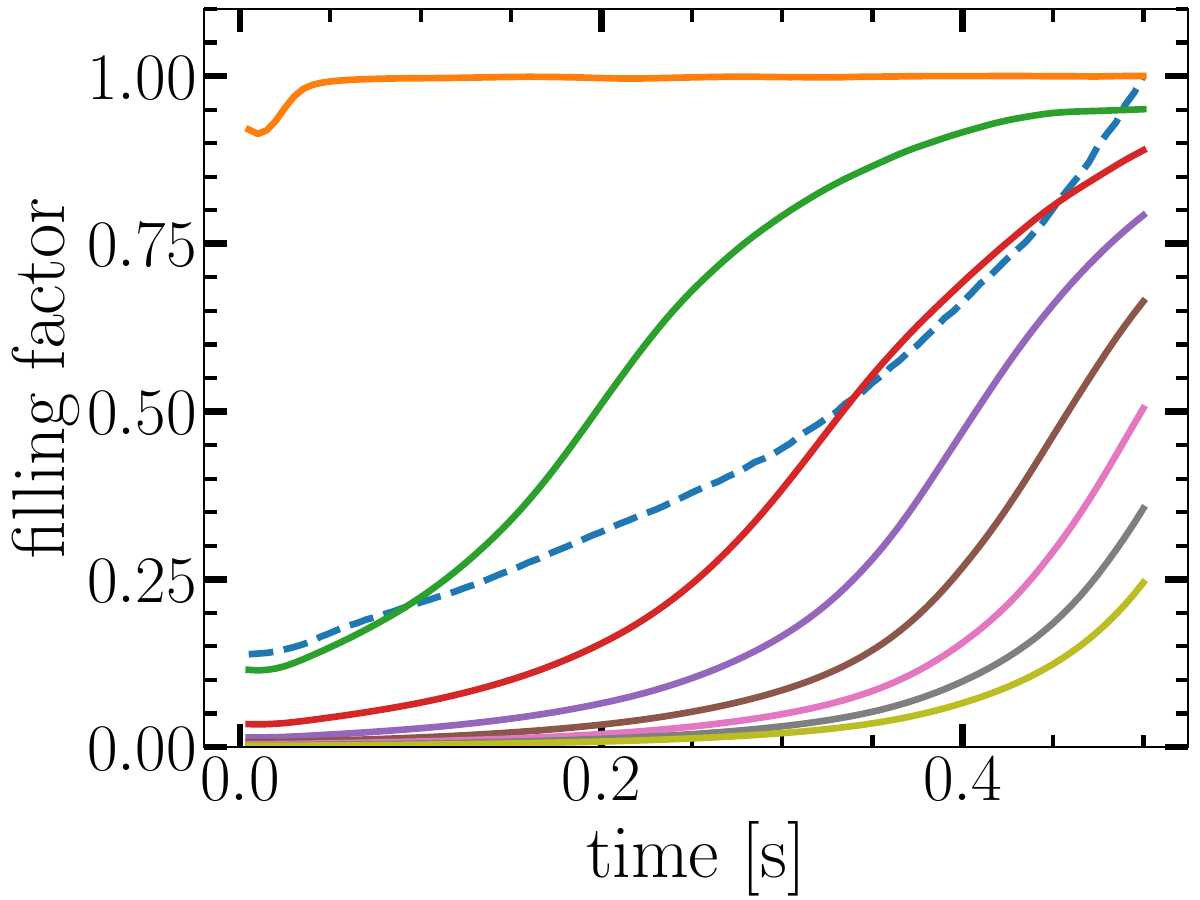}}
    \caption{Filling factors as a function of time. Shown is the enclosed burnt volume fraction of eight spheres around the center with radii from 100 to 800 km in jumps of 100 km (orange to gold). Overplotted is the deflagration front distance from the center (in blue). }
    \label{fig:fill_prof_t}
\end{figure*}

{ To quantify the efficiency of burning, angle-averaged quantities are shown as a function of time. Figure~\ref{fig:fill_prof_t}} shows the volume occupied by completely burned material in enclosed spheres around the center divided by the enclosed volume of the spheres, referred to as the filling factor. We show these filling factors as a function of time within eight different spheres ranging from 100 km (the initial radius of ignition) to 800 km for our six simulations (Table~\ref{tab:prop}). We also plot the maximal distance of the deflagration front from the center ($R_{\rm max}$).
\begin{figure*}
\centering  
    \includegraphics[scale=0.38]{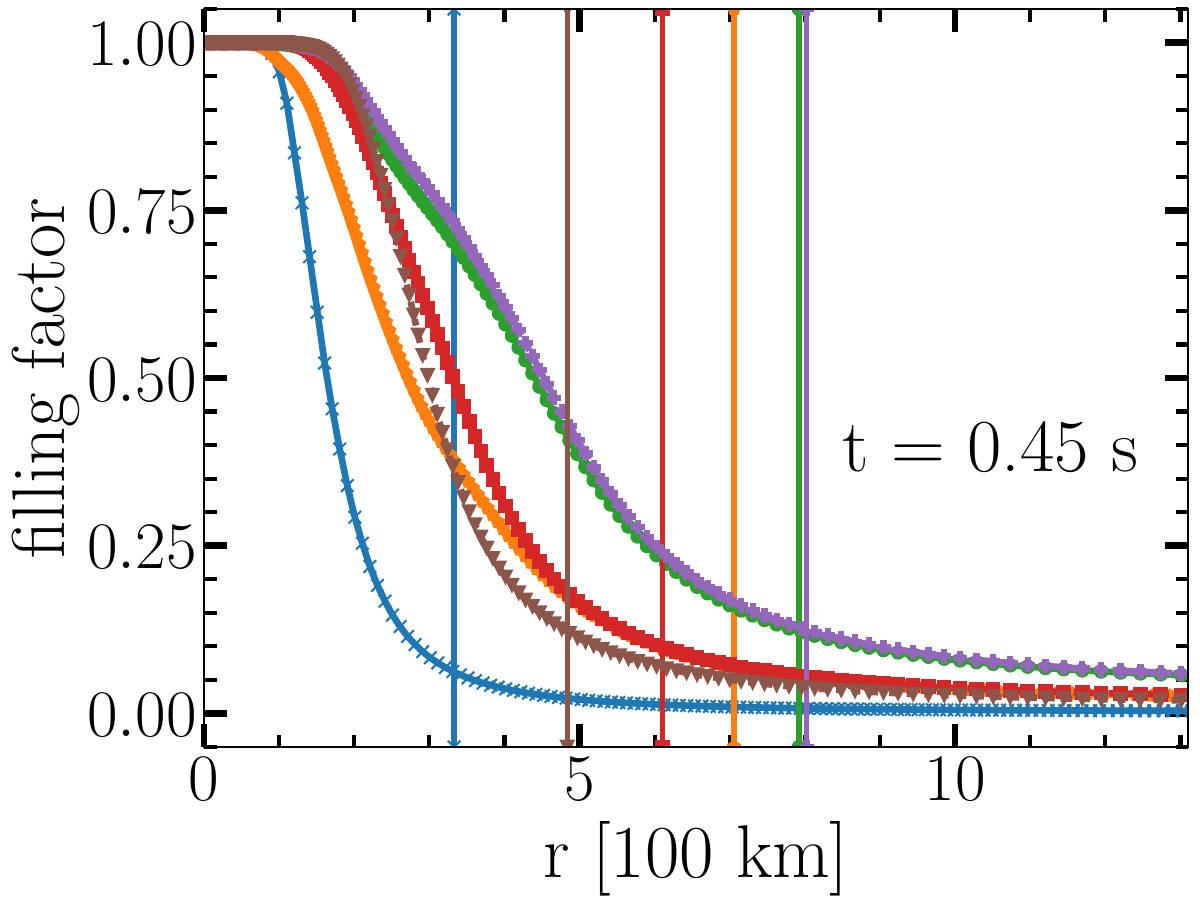}
    \includegraphics[scale=0.38]{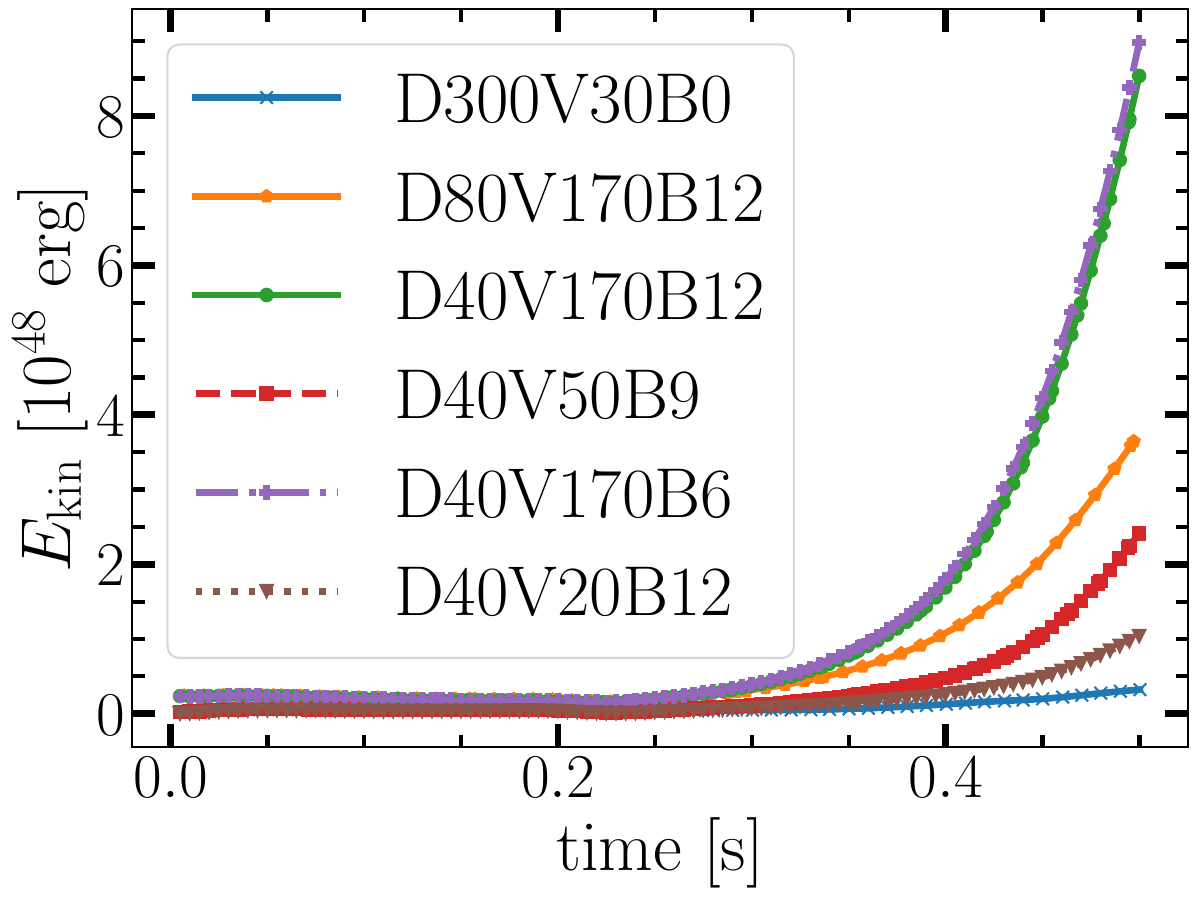}
    \caption{{Left: filling factor as a function of radius of our simulations at $t=0.45~{\rm s}$. The vertical lines denote the distance of the deflagration front from the center. Right: kinetic energy as a function of time.}}
    \label{fig:fill_ek_t}
\end{figure*}

In pure hydrodynamic simulations or in simulations with large-scale turbulent fields {(D300V30B0; panel (a))}, the filling factor for most radii is smaller than $\approx 10\% $, and even in the central $200~{\rm km}$ it reaches only 0.35. The burning is not complete and far away from spherical deflagration models in which the filling factor is exactly 1 inside the deflagration front. 

Turbulent fields with eddies that are similar in size to the pockets or slightly larger result in a drag of the unburned material, effectively increasing the filling factors (D80V170B12). Small-scale turbulence fields with eddy sizes that are smaller than the pockets
result in efficient mixing of the plumes with the fuel, burning away the pockets. This leads to efficiencies that exceed $75 \%$ in the inner 400 km of the WD. The burning in this simulation is much closer to a spherical model. 

{Figure~\ref{fig:fill_ek_t} shows the filling factor as a function of the radius from the center at $t=0.45~{\rm s}$ (a), and the total kinetic energy as a function of time (b). Strong small-scale turbulent fields (D40V170B12, D40V170B6) result in a large effective radial burning speed, faster increase in kinetic energy, and, as discussed above, large filling factors.}
{
For models with small-scale strong turbulence, the higher efficiency in burning will shorten the time until the preexpansion of the WD sets in, resulting in a shorter phase of high-density nuclear burning.
}

%% file: conclusions.tex
We presented a series of simulations demonstrating that preexisting small-scale turbulent fields result in a qualitative change in the explosion of near-$M_{{\rm Ch}}$ mass WDs. 
We examined various turbulence conditions with different diffusion radii ($r_{\rm D}=40,~80,~300~{\rm km}$), turbulent velocities {($v_{\rm rms}=20,~30,~50,~170~{\rm km/s}$),} and turbulent magnetic field strengths {($B_{\rm rms}=0,~10^{6},~10^{9},~10^{12}~{\rm G}$)} (see Table~\ref{tab:prop} and Figure~\ref{fig:turb_setup} for a snapshot of a simulation with small-scale strong turbulence fields D40V170B12 at $t=0~{\rm s}$). The turbulence field drags the burned material into unburned pockets and entrains the deflagration front, effectively reducing the distance between the burned and unburned components, such that the laminar front is able to burn the fuel (Figures~\ref{fig:comp_models_flam_xy}-\ref{fig:comp_models_t045_temp}).  This qualitative change serves as a first step toward a solution to the long-standing problem of insufficient expansion of the WD due to incomplete burning during the deflagration phase, as found in previous multidimensional studies (see Section~\ref{sec:intro}). {Although magnetic fields may have a similar effect on entrainment, we show that the preexisting turbulence dominates over magnetic fields that are less than a percent of the equipartition field. However, magnetic fields still slightly reduce the formation of small structures (Figures~\ref{fig:comp_models_flam_xy}~and~\ref{fig:comp_models_t045_temp})}.  

Overall, based on our results, we conclude the following:
\begin{itemize}
\item 
turbulent fields have a significant impact on the early development of the flame. 
We note that preexisting turbulence must be expected from the preexplosive burning (see Section~\ref{sec:intro} \citealt{HoeflichStein2002}; \citealt{zingale11}). {Large $B$ fields smooth out small scales.} Evidence for high-magnetic fields is related to late-time LCs and spectra (see Section~\ref{sec:magnetic} and \citealt{Hristovetal2021}).

\item 
in pure hydrodynamic simulations or simulations with large-scale turbulent fields, such as in \cite{Gamezoetal2003} and simulation D300V30B0 of this study, the early burning phase can be characterized by large pockets of unburned material. The burned material consists at most of only $\approx 10\% $ of the volume (panel (a) of Figure~\ref{fig:fill_prof_t}). As a consequence, the burning energy mainly drives the rise of the burned plumes rather than the expansion of the WD (panel (b) of Figure~\ref{fig:fill_ek_t}).

\item 
turbulent velocity fields with eddies that are comparable in size to the pockets or slightly larger result in a drag of the unburned material (passive flow), effectively increasing the burning efficiency, but only to a level insufficient for a considerable expansion of the WD. (left column of Figures~\ref{fig:comp_models_flam_xy}~and~\ref{fig:comp_models_t045_temp} and Figure~\ref{fig:fill_prof_t}~panel~(b)).

\item 
small-scale turbulence fields with eddy sizes that are smaller than the pockets
result in an efficient mixing of the fuel into the plumes, burning the pockets away. This leads to efficiencies that exceed $75 \%$ in the inner 400 km of the WD, and a continuous increase of $V_{{\rm eff}}$. As a sequence, these 3D simulations approach to the evolution of spherical simulations resulting in a very effective expansion of the WD (Figure~\ref{fig:fill_ek_t} panel (b)).
\end{itemize}

We must stress that although our high-resolution simulation D40V170B12 yielded the most efficient burning, resolution alone cannot cause this qualitative difference. This has been shown (Poludnenko, private communication) through simulations with increasing resolution reaching a finer resolution than our simulations. {It is also supported by simulations D40V50B9 and D40V20B12, which have the same resolution as D40V170B12, but weaker turbulence, and resulted in less preexpansion.}
The crucial ingredient is the presence of turbulent fields, in particular, turbulence with sufficiently small eddies. {Note, though, that with increasing resolution the dissipation of magnetic fields might become increasingly important.} 

Finally, we want to emphasize the limitations of the current study. 
Here, we only considered central ignition. However, in a corresponding ongoing study, we investigate the role of preexisting fields on deflagration that starts off-center.
In addition, in the present study we focus on the early deflagration phase up to 0.5 s. In a future study, we will simulate the entire deflagration phase. 
This will require the removal of our constraint of zeroing the velocities at radius greater than 1,500 km, and the inclusion of quasi-NSE and carbon burning. Varying the deflagration-to-detonation transition will allow us to qualitatively compare the models to a wide variety of SN observations, including a direct
comparison with the production of EC elements (see Section~\ref{sec:intro}).

Although we showed that strong small-scale turbulent fields are likely essential for 
the overall evolution of the explosion and pave the way for the future, more detailed MHD studies of the initial conditions are needed.
For example, our initial WD has a central density of $\approx 10^9~ {\rm g~cm^{-3}}$, a value typically found in 
observations (Section~\ref{sec:intro}). In future studies, we will focus on WDs with higher central densities. Such high densities are expected to produce more EC elements, and therefore it would be specifically interesting to investigate the effects of turbulence fields on the production of EC elements in these cases. 

{As discussed in Section~\ref{sec:magnetic},
an effective dynamo is required to amplify the magnetic fields beyond $10^6~{\rm G}$ indicated from observations. This amplification may proceed until $B$ approaches the equipartition values. However, our study is limited to magnetic fields less than a few percent of the saturation because, for higher $B$, the morphology will not necessarily resemble a field produced by hydrodynamics. It warrants further investigation whether ultrastrong $B$ fields, closer to equipartition values, will dominate the burning rate over the turbulence in the early deflagration phase.}